\documentclass[traditabstract]{aa}
\usepackage{amssymb}
\usepackage{graphics}
\usepackage{tabularx}
\usepackage{color}
\usepackage{natbib}
\usepackage{rotating}

\def\az{2013\,AZ\ensuremath{_{60}}}
\def\dr{2012\,DR\ensuremath{_{30}}}
\def\gz{2002\,GZ\ensuremath{_{32}}}
\def\tiunit{{\rm J}{\rm m}^{-2}{\rm K}^{-1}{\rm s}^{-1/2}}

\def\refmark#1{\textbf{#1}}

\begin{document}


\title{Physical properties of the extreme centaur and super-comet candidate \az{}}

\author{%
A. P\'al\inst{1,2} \and
Cs. Kiss\inst{1} \and
J. Horner\inst{3,4} \and
R. Szak\'ats\inst{1} \and
E. Vilenius\inst{5,6} \and
Th. G. M\"uller\inst{5} \and
J. Acosta-Pulido\inst{7,8} \and
J. Licandro\inst{7,8} \and
A. Cabrera-Lavers\inst{7,8} \and
K. S\'arneczky\inst{1} \and
Gy. M. Szab\'o\inst{9,1} \and
A. Thirouin\inst{10} \and
B. Sip\H{o}cz\inst{11} \and
\'A. D\'ozsa\inst{9} \and
R. Duffard\inst{12} 
}

\institute{%
Konkoly Observatory, MTA Research Centre for Astronomy and Earth Sciences,
	Konkoly-Thege Mikl\'os \'ut 15-17,
	1121 Budapest, Hungary; e-mail: \texttt{apal@szofi.net} \and
Department of Astronomy, Lor\'and E\"otv\"os University, 
	P\'azm\'any P\'eter s\'et\'any 1/A, 
	1117 Budapest, Hungary \and
Computational Engineering and Science Research Centre, 
	University of Southern Queensland, Toowoomba, 
	Queensland 4350, Australia \and
Australian Centre for Astrobiology, UNSW Australia, 
	Sydney, New South Wales 2052, Australia \and
Max-Planck-Institut f\"ur extraterrestrische Physik, 
	Postfach 1312, Giessenbachstr., 85741 Garching, Germany \and
Max-Planck-Institut f\"ur Sonnensystemforschung, 
	Justus-von-Liebig-Weg 3, 37077 G\"ottingen, Germany \and
Instituto de Astrof\'{\i}sica de Canarias, 
	38205 La Laguna, Tenerife, Spain \and
Departamento de Astrosf\'{\i}sica, Universidad de La Laguna, 
	38206 La Laguna, Tenerife, Spain \and
Gothard Astrophysical Observatory, 
	Lor\'and E\"otv\"os University, 
	9700 Szombathely, Hungary \and
Lowell Observatory,
	1400 W Mars Hill Rd, 86001, Arizona, USA \and
Centre for Astrophysics Research, 
	University of Hertfordshire, 
	College	Lane, Hatfield AL10 9AB, UK \and
Instituto de Astrof\'{\i}sica de Andaluc\'{\i}a - CSIC, 
	Apt 3004, 18008 Granada, Spain 
}


\date{Received \dots; accepted \dots}

\abstract{%
We present estimates of the basic physical properties
-- including size and albedo -- of the extreme Centaur
\az{}. These properties have been derived from optical and thermal infrared
measurements. Our optical measurements revealed a likely full period 
of $\approx 9.4\,{\rm h}$ with a shallow amplitude of 4.5\%. By combining
optical brightness information and thermal emission data, we are able
to derive a diameter of $62.3\pm5.3\,{\rm km}$ and a 
geometric albedo of 2.9\% -- corresponding to an extremely dark
surface. Additionally, our finding of $\gtrsim50\,\tiunit$ for
the thermal inertia is also noticeably for objects in such a distance.
The results of dynamical simulations yield an unstable orbit,
with a 50\% probability that the target will be ejected from the Solar System
within 700,000\,years. The current orbit of this object as well as its 
instability could imply a pristine cometary surface. 
This possibility is in agreement with the observed low geometric albedo
and red photometric colour indices for the object, which are a good
match for the surface of a dormant comet -- as would be expected for a
long-period cometary body approaching perihelion. Despite the fact it
was approaching ever closer to the Sun, however, the object exhibited
star-like profiles in each of our observations, lacking any sign of
cometary activity. By the albedo, 
\az{} is a candidate for the darkest body among the known TNOs.}


\maketitle


\section{Introduction}

The object \az{} is a recently discovered extreme Centaur, moving
on an eccentric orbit with $e\approx 0.992$ and a perihelion distance of
$q\approx 7.9\,{\rm AU}$. As a result, \az{} is among the TNOs with the 
largest known aphelion distance at $\approx 1950\,{\rm AU}$.
\az{} may be classified as a Centaur, based on its perihelion
distance \citep{horner2003}. However, due to its large semimajor
axis, it could equally be considered to be a scattered disk object 
\citep{gladman2008}. Its Tisserand parameter \citep{duncan2004} w.r.t. Jupiter 
is $T_{\rm J}=3.47$ which is typical for Centaurs \citep{horner2004a,horner2004b}
and differs from that of Jupiter family comets ($2<T_{\rm J}<3$) and 
especially for from that of Damocolids and Halley-type comets 
\citep[$T_{\rm J}<2$, see][]{jewitt2005} that exhibit cometary dynamics. 

In order to recover the basic physical and surface characteristics of this 
object, we need measurements both in the visual and in the thermal infrared 
range. Optical data can yield information about the intrinsic colours,
the absolute brightness, rotational period, shape and surface 
homogeneity of the object, while thermal observations aid us to 
decide whether we see a ``large but dim''
or a ``small but bright'' surface. For this latter purpose, 
Herschel Space Observatory \citep{pilbratt2010} is an ideal instrument
since the expected peak of the thermal emission is close to the shortest
wavelengths of its PACS detector \citep{poglitsch2010}. 

In our current analysis, we follow the same methodology as presented in
our previous study of the Centaur \dr{} \citep{kiss2013}, another 
object moving on a similar orbit to \az{}.
The structure of this paper is as follows. 
In Sec.~\ref{sec:observations}, we describe our observations, 
including the detection of thermal emission by Herschel/PACS, optical 
photometry by the IAC-80 telescope (Teide Observatory, Tenerife, Spain), 
optical reflectance by the Gran Telescopio Canarias (GTC, Roque de los 
Muchachos Observatory, La Palma, Spain) and near-infrared photometry by 
the William Herschel Telescope (WHT, Roque de los Muchachos Observatory, 
La Palma, Spain).
In Sec.~\ref{sec:thermal}, we derive the basic physical 
properties of the object by applying well understood thermophysical models.
The dynamics of \az{} are then discussed in Sec.~\ref{sec:dynamics}. 
Finally, our results are summarized in Sec.~\ref{sec:discussion}.


\begin{table*}
\caption{Summary of Herschel observations of \az{}, obtained in the 
DDT program {\it DDT\_ckiss\_3}. The columns are: 
i) visit; 
ii) observation identifier; 
iii) date and time;
iv) duration; 
v) filters configuration;
vi) scan angle direction with respect to the detector array.}
\label{table:herschelobs}
\begin{center}\begin{tabular}{cccccr}
\hline
Visit   & OBSID & Date \& Time  & Duration & Filters          & Scan angle \\
        &       &   (UT)        & (s)      & ($\mu$m/$\mu$m)  & (deg) \\            
\hline
        & 1342268974 & 	2013-03-31 18:10:51 &	1132 &	70/160	&	70	\\
Visit-1 & 1342268975 & 	2013-03-31 18:30:46 &  	1132 &  70/160	&	110	\\
        & 1342268976 & 	2013-03-31 18:50:41 &  	1132 & 	100/160	&	70	\\
        & 1342268977 & 	2013-03-31 19:10:36 &  	1132 &	100/160	&	110	\\
\hline
        & 1342268990 & 	2013-03-31 23:46:55 &  	1132 &	70/160	&	110	\\	
Visit-2 & 1342268991 & 	2013-04-01 00:06:50 &  	1132 &	70/160	&	70	\\	
        & 1342268992 & 	2013-04-01 00:26:45 &  	1132 &	100/160	&	110	 \\	
        & 1342268993 & 	2013-04-01 00:46:40 & 	1132 &	100/160	&	70	 \\	
\hline
\end{tabular}\end{center}\vspace*{-3mm}
\end{table*}

\section{Observations and data reduction}
\label{sec:observations}

\subsection{Thermal observations and flux estimations}
\label{sec:thermalobservations}

Thermal infrared images have been acquired with the 
Photoconductor Array Camera and Spectrometer \citep[PACS][]{poglitsch2010}
camera of the Herschel Space Observatory \citep{pilbratt2010} in
two series, each of 1.3\,hour duration. As we summarize in Table.~\ref{table:herschelobs},
these two measurement cycles were separated by more than 4 hours, allowing the 
target object \az{} to move, but still be in the same field of view. 
This type of data collection has almost exclusively been employed 
in the ``TNOs are Cool!'' Open Time Key Programme of Herschel
\citep{mueller2009,mueller2010,vilenius2012}. For
both series of measurements, we used both the blue/red 
($70/160\,\mu{\rm m}$) and green/red ($100/160\,\mu{\rm m}$) 
channel combinations. This scheme allowed us to use the second 
series of images as a background for the first series
of images (and vice versa) in order to eliminate the systematic effects
of the strong thermal background. This type of data acquisition and
the respective reduction scheme were described in our former works 
related to both the ``TNOs are Cool!'' project \citep[see e.g.][]{vilenius2012,pal2012sedna} 
and subsequent measurements \citep[see e.g.][]{kiss2013}. 

Unfortunately, the astrometric uncertainties of \az{} were relatively
large at the time of Herschel observations, due to its rather recent discovery.
Thus, the apparent position of the object was slightly 
($\approx 29^{\prime\prime}$) off from the image
center, which also implied that the double-differential photometric 
method \citep{kiss2014} yielded larger photometric uncertainties. 
In addition, shortly before the Herschel observations, on February 
16, 2013 (at OD-1375\footnote{http://herschel.esac.esa.int/Docs/Herschel/\\/html/ch03s02.html\#sec3:DeadMat}), 
one half of the red ($160\,\mu{\rm m}$) channel pixel array became 
faulty. Hence, only the images from the first visit were sufficient to
obtain fluxes at $160\,\mu{\rm m}$ and it was not possible to create
double-differential maps in this channel. 

\begin{figure}
\begin{center}
\begin{tabular}{cccc}
& $70\,\mu{\rm m}$ & $100\,\mu{\rm m}$ & $160\,\mu{\rm m}$ \\
\begin{sideways}\hspace*{10mm}Individual\end{sideways} & 
\hspace*{-4mm}\resizebox{30mm}{!}{\includegraphics{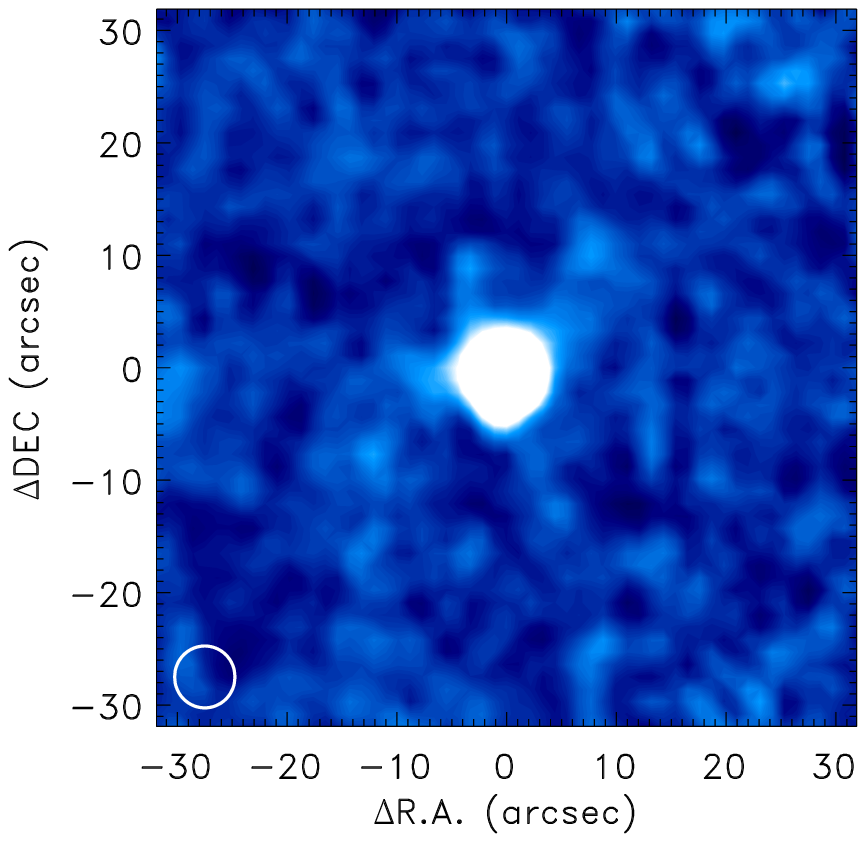}} \hspace*{-4mm}&
\hspace*{-4mm}\resizebox{30mm}{!}{\includegraphics{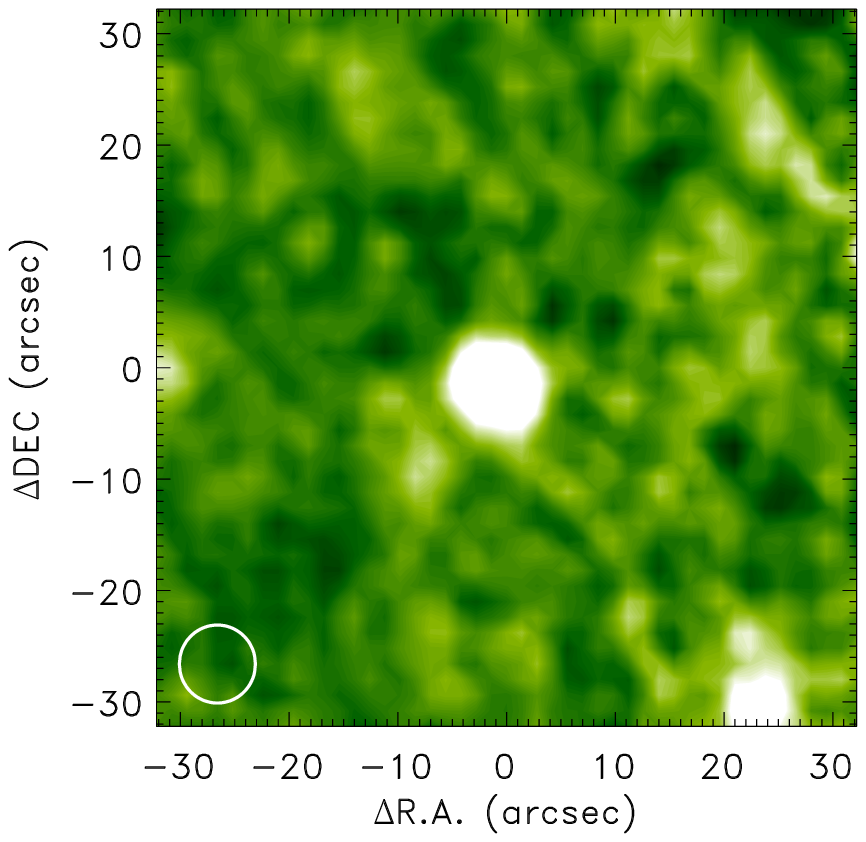}}\hspace*{-4mm} &
\hspace*{-4mm}\resizebox{30mm}{!}{\includegraphics{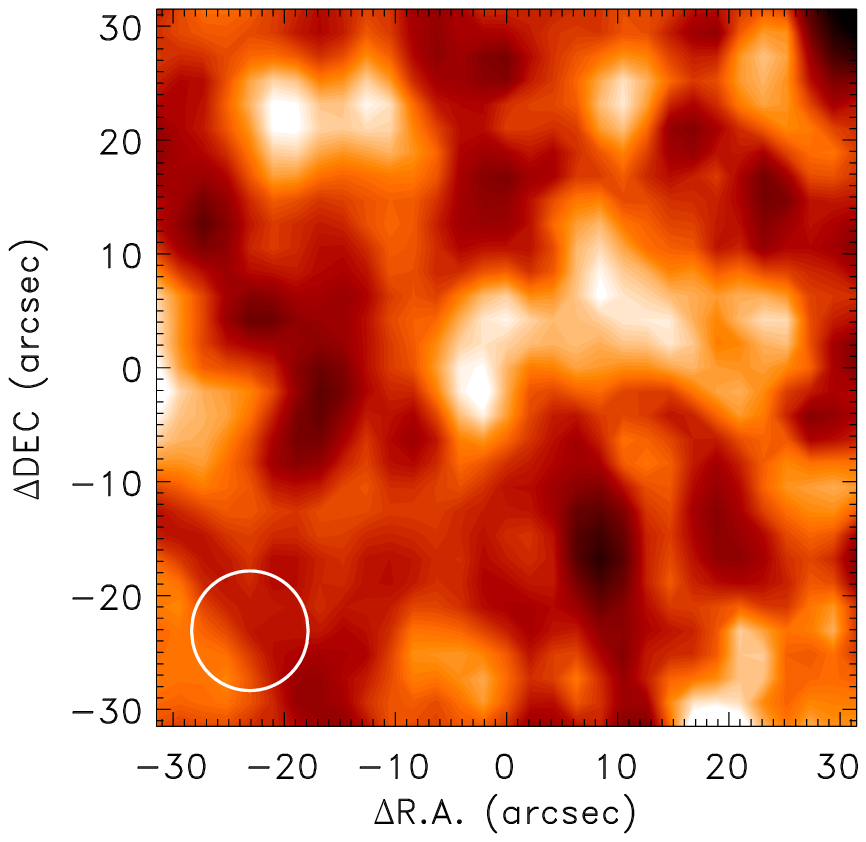}} \\[-6mm]
\begin{sideways}\hspace*{10mm}Double differential\end{sideways} & 
\hspace*{-4mm}\resizebox{30mm}{!}{\includegraphics{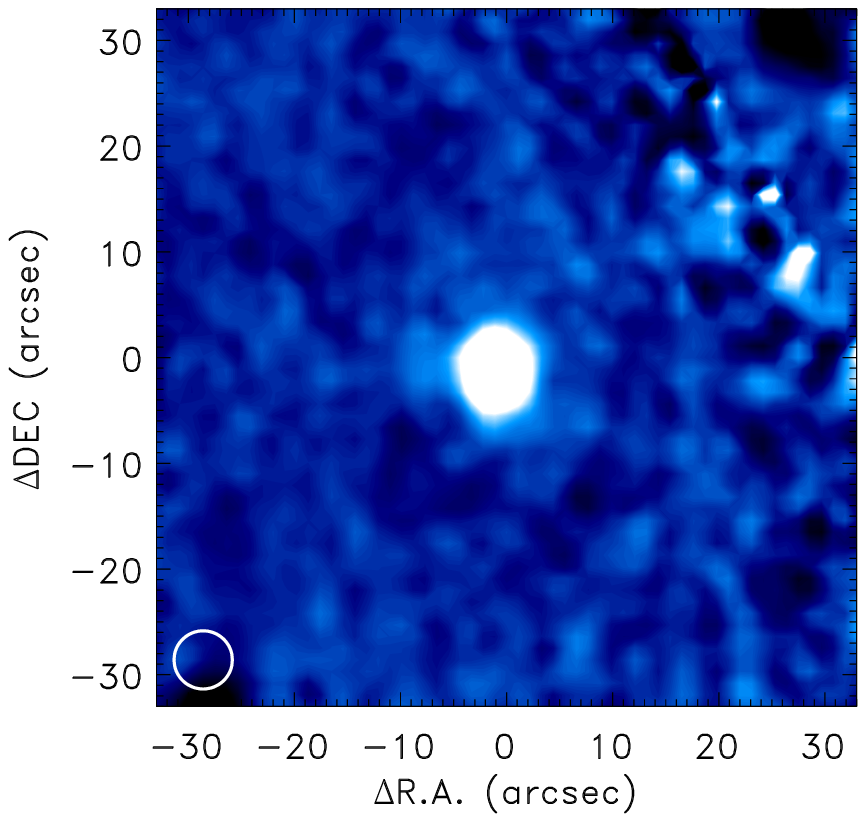}} \hspace*{-4mm}&
\hspace*{-4mm}\resizebox{30mm}{!}{\includegraphics{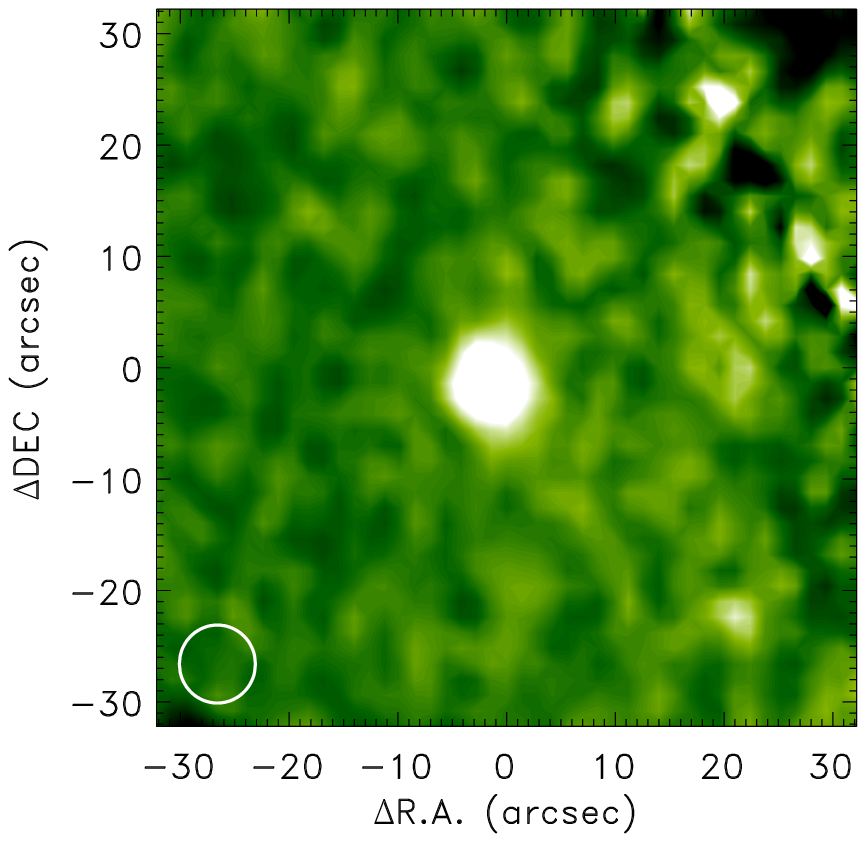}}\hspace*{-4mm} &
\end{tabular}
\end{center}\vspace*{-4mm}
\caption{Image stamps showing the Herschel/PACS maps 
of \az{} in the $70\,\mu{\rm m}$ (blue),
$100\,\mu{\rm m}$ (green), and $160\,\mu{\rm m}$ (red) channels.
Each stamp covers an area of $64^{\prime\prime}\times64^{\prime\prime}$, while 
the tick marks on the axes show the relative positions in pixels. 
The effective beam size (i.e. the circle with a diameter corresponding 
to the full width at half maximum) is also displayed in the lower-left 
corners of the stamps. The upper panels show the stamps directly combined from
the individual frames where the lower stamps are obtained using the double-differential
method. Due to the failure of the half of the red channel and the astrometric
uncertainties, the second visit is nearly unusable in red, hence double-differential
maps cannot be created.}
\label{fig:azimagestamps}
\end{figure}

Raw Herschel/PACS data has been processed in the HIPE environment
\citep{ott2010} with custom scripts described in \cite{kiss2014}.
The double-differential maps were created and analyzed 
using the FITSH package \citep{pal2012fitsh}. The resulting images
are displayed in Fig.~\ref{fig:azimagestamps}.
Photometry on the individual
as well as on the combined double-differential images were performed 
by using aperture photometry where the fluxes were corrected by
the respective growth curve functions. Photometric uncertainties were 
estimated by involving artificial source implantation in a Monte-Carlo
fashion \citep{santossanz2012,mommert2012,kiss2014}. This method works
both for the double-differential images (blue and green channels)
as well as on the individual maps (here, the red channel).

Based on the individual images, we obtained thermal fluxes of
$36.6\pm2.9$, $25.2\pm3.7$ and $15.9\pm4.5\,{\rm mJy}$ in the blue,
green and red wavelengths, respectively. By involving the double-differential
maps, we derived $32.5\pm2.2$ and $23.0\pm2.8\,{\rm mJy}$ in the 
blue and green regimes. Due to the lower level of confusion noise 
(see Fig.~\ref{fig:azimagestamps}), the accuracy of the latter series
of fluxes is better. Therefore, for further modelling we adopt the 
double-differential fluxes for blue and green. Thermal fluxes should 
undergo colour correction according to the temperatures of the bodies
\citep[see][for the respective coefficients]{poglitsch2010}.
Since the subsolar temperature of \az{} is around $110\,{\rm K}$, the colour 
correction is negligible (less than a percent) in blue and green while it 
is +4\% in red. Our reported fluxes consider the respective colour 
correction factors. In addition, in the error estimation of the aforementioned 
fluxes, we included the 5\% systematic error for the absolute flux calibration 
as well \citep{balog2014}. 
The summary of these thermal fluxes are reported in Table~\ref{table:herscheldata}.

\begin{table}
\caption{Thermal fluxes 
of \az{} derived from our Herschel measurements.} 
\label{table:herscheldata}
\begin{center}\begin{tabular}{llrr}
\hline
Band & $\lambda$ & Flux \\
\hline
B	& $70\,{\rm\mu m}$	& $32.5\pm2.2\,{\rm mJy}$	\\
G	& $100\,{\rm\mu m}$	& $23.0\pm2.8\,{\rm mJy}$	\\
R	& $160\,{\rm\mu m}$	& $15.9\pm4.5\,{\rm mJy}$	\\
\hline
\end{tabular}\end{center}\vspace*{-3mm}
\end{table}

\begin{figure}
\begin{center}
\resizebox{85mm}{!}{\includegraphics{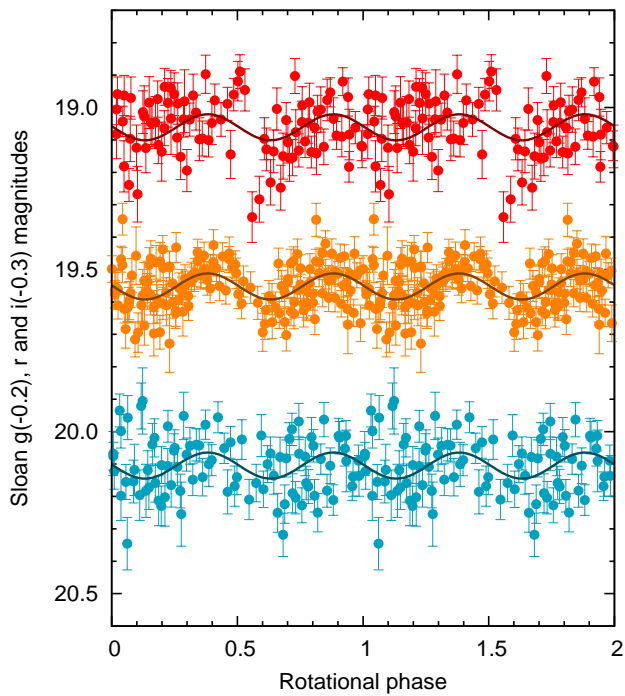}}
\end{center}\vspace*{-4mm}
\caption{Folded optical light curves
of \az{} using photometric data taken on 6 subsequent nights of
2013 November 4 -- 9. Note that the folding frequency is related to the 
preferred double-peaked solution, $n=(5.11/2)\,{\rm d}^{-1}$.
See text for further details. }
\label{fig:azlightcurve}
\end{figure}

\begin{figure}
\begin{center}
\resizebox{85mm}{!}{\includegraphics{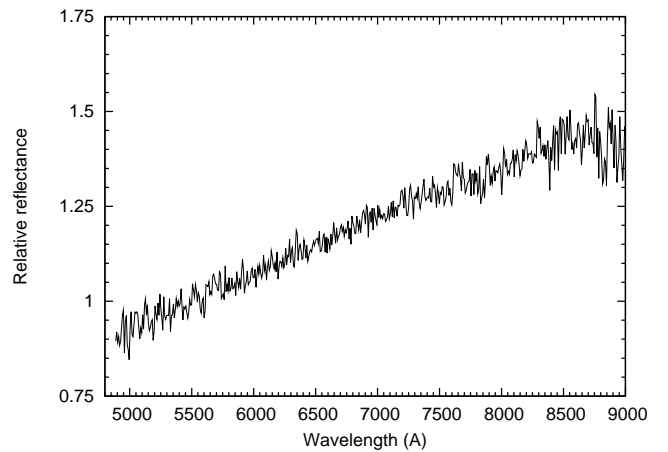}}
\end{center}\vspace*{-4mm}
\caption{Reflectance spectrum of
\az{}, taken with the OSIRIS spectrometer on the GTC in January, 2014.
This spectrum is normalized to unity at $\lambda=5500\,{\rm \AA}$.}
\label{fig:reflectance}
\end{figure}


\subsection{Optical photometry}
\label{sec:photometric}

Since \az{} has recently been discovered, one of our goals was to
obtain precise photometric time series for this object in order 
to estimate both the absolute magnitudes (in various passbands) and
the rotational period based on light curve variability. For these
purposes, we used the IAC-80 telescope located at Teide Observatory,
Tenerife. During our observing runs, we used the CAMELOT
camera, equipped with a CCD-E2V detector of $2{\rm k}\times 2{\rm k}$, 
with a pixel scale of $\approx 0.3^{\prime\prime}$ 
providing a field of view of roughly $10^{\prime}\times 10^{\prime}$.
Time series were gathered for several hours on the nights of 2013 November
4 -- 9 using Sloan $g'$, $r'$ and $i'$ filter sets. Since \az{} can 
currently be found at the edge of the spring Sloan field, several dozen stars 
with accurate reference magnitudes were available on each image. 
The night conditions were photometric on 5th, 7th and 9th of 2013 November
where the individual photometric uncertainties were nearly constant and varied
between $0.04$ and $0.05$\,mags in Sloan $r'$ band. The conditions were 
worse on the three another nights (4th, 6th and 8th) when the formal
uncertainties scattered in the range of $0.04-0.08$ indicating the variable
transparency of the sky (which was also notable during the observations).
The worst conditions were on the first night where some of the measurements 
had a formal uncertainty of $0.09$.

The scientific images were analyzed with the standard calibration,
source extraction, astrometry, cross matching and photometry tasks
of the FITSH package \citep{pal2012fitsh}. As a hint, we used the MPC predictions
for the target coordinates and then we performed individual centroid fit
on each image and smoothed with polynomial regression for a better (much more 
precise and accurate) input for aperture photometry. Instrumental 
magnitudes were then extracted for both the reference stars and the
target itself and then after applying the standard photometric transformations,
we obtained the intrinsic $g'$, $r'$ and $i'$ magnitudes. 

We searched for possible light curve variations using the most frequently
sampled Sloan r' band data (every second image was taken in Sloan $r'$
while every fourth image was in $g'$ and $i'$). In order to search for
periodic variations in our data set, we fitted a function in a form of
$f(t)=a+b\cdot\sin(2\pi nt)+c\cdot\cos(2\pi nt)$ to the Sloan $r'$ photometric series where $t$ indicates
the independent value (the time in this case). 
If the value of $n$ is scanned
in the appropriate domain ($n=0.01\dots15$) with a proper stepsize ($n=0.01$,
that is $\sim 8$ times smaller than the stepsize implied
by the Nyquist criterion), then the parameters
$a$, $b$ and $c$ can be obtained via a simple weighted linear least squares 
fit procedure. The unbiased $\chi^2$ values can then be compared with the
reference value of $\chi^2_0$. This reference value is obtained 
when $n$ is set to $0$ and the error bars are scaled by a factor of $1.51$ 
to yield a $\chi^2_0$ equivalent to the degrees of freedom.
The difference between the $\chi^2_0$ and the $\chi^2$ related to the 
adopted period tells the significance of the detection while various additional
possible periods can also be checked and/or ruled out according to 
the difference between the respective $\chi^2$ values. 
We found a significant variation ($\chi^2_0-\chi^2=24.2$) with a corresponding amplitude 
of $\Delta{\rm r'}=0.045\pm0.007$ that
has a frequency of $n=5.11\pm0.12\,{\rm d}^{-1}$. The folded 
light curves are displayed in Fig.~\ref{fig:azlightcurve}. The mean magnitudes 
of these observations were 
$g'_1=20.274\pm0.013$, $r'_1=19.519\pm0.009$ and $i'_1=19.316\pm0.013$.

We have to note here that due to the daily aliases, the peaks around 
$n\pm 1\,{\rm d}^{-1}$ are also remarkable and there is a non-negligible
chance that one of these frequencies belong to the intrinsic rotation
of the object. The peak at $n=6.11$ has a $\chi^2$ value which is only larger
than that of the main peak by $3.5$. 

In general, minor bodies in the Solar System feature double-peaked
light curves. Hence, the rotational frequency of \az{} is more
likely $n_{\rm rot}=n/2\,{\rm d}^{-1}$, equivalent to a period of 
$P_{\rm rot}=9.39\pm0.22\,{\rm h}$. In order to test the significance of a
double-peaked light curve solution, we coadded a sinusoidal component 
with half of the frequency to the primary variations. The amplitude of
this component is found to be $0.013\pm0.008\,{\rm mag}$. This is only
a 1.7-$\sigma$ detection, however, a good argument for confirming the
assumption for an intrinsic rotation period 
of $P_{\rm rot}\approx 9.4\,{\rm h}$. 

In addition, we repeated the photometric observations for \az{} in 
2014 January 28 in $g'$ and $r'$ bands. The results of these
photometric measurements yielded the Sloan magnitudes of 
$g'_2=19.71\pm0.04$ and $r'_2=18.99\pm0.03$. During the first series
of measurements (in 2013 November), the geocentric and heliocentric
distance of \az{} were $\Delta_1=8.176\,{\rm AU}$ and $r_1=8.244\,{\rm AU}$,
respectively, while in 2014 January 28, these distances were 
$\Delta_2=7.148\,{\rm AU}$ and $r_2=8.114\,{\rm AU}$. Based on these distances,
the expected change in the apparent brightness was 
$5[\log_{10}(r_2\Delta_2)-\log_{10}(r_1\Delta_1)]=-0.326$,
however, the actual brightness changes were $\Delta g'=-0.56\pm0.04$
and $\Delta r'=-0.53\pm0.03$, whose mean is $\Delta m=-0.54\pm0.03$.  
Since the phase angle of \az{} was $\alpha_1=6.9^\circ$
in 2013 November 5 and $\alpha_2=1.5^\circ$ in 2014 January 28, these values
imply a phase correction factor of 
$\beta=[(0.54\pm0.03)-0.326]/(6.9-1.5)=0.039\pm0.006\,{\rm mag}/{\rm deg}$.
This is is rather good agreement with MPC observations. Based on the 
MPC observation database, the best-fit phase correction parameter can also
be derived, however, with a larger uncertainty: 
$\beta_{\rm MPC}=0.040\pm0.025\,{\rm mag}/{\rm deg}$.

These parameters allowed us to derive the absolute brightness of the
object \az{} in a manner described in the following. First, 
we employed simple Monte-Carlo run whose input were the observed Sloan 
brightnesses, the derived phase correction factor as well as the 
parameters and the respective uncertainties of 
the corresponding Sloan-UBVRI transformation equation 
\citep[for converting $g'$ and $r'$ brightnesses to $V$, see][]{jester2005}.
This Monte-Carlo run yielded a value of $H_V=10.42\pm0.07$. Next, we checked
the available photometric data series presented in the MPC database which yielded
slightly fainter values, namely $H_{V,{\rm MPC}}=10.60\pm0.15$. In order
to reflect MPC photometry in our derived absolute brightness value, 
we adopted the weighted mean value of these two values, namely $H_V=10.45$ 
with a conservative uncertainty of $\pm0.10$ in the subsequent modelling.

\begin{figure*}
\begin{center}
\resizebox{!}{46mm}{\includegraphics{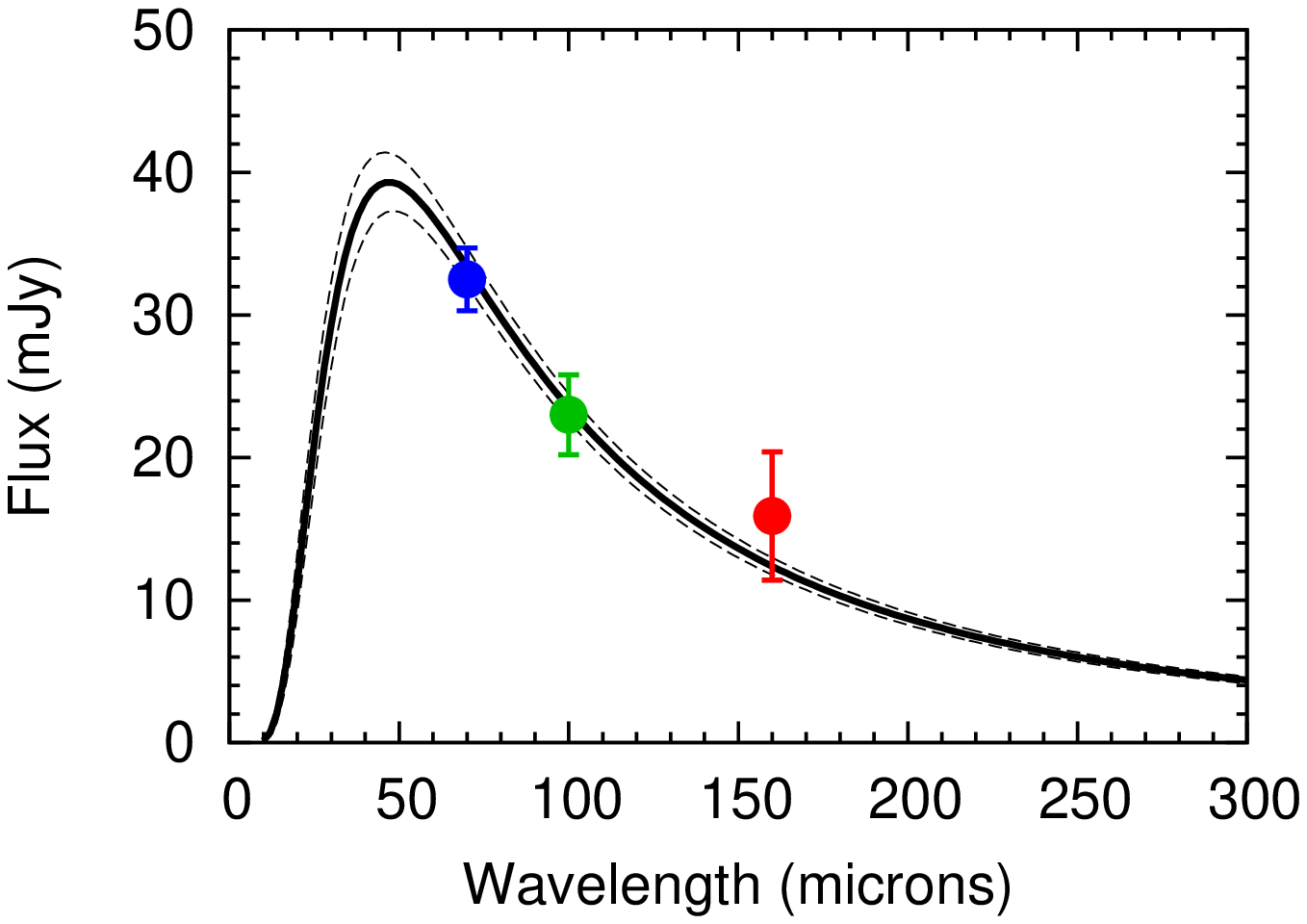}}%
\resizebox{!}{46mm}{\includegraphics[angle=90]{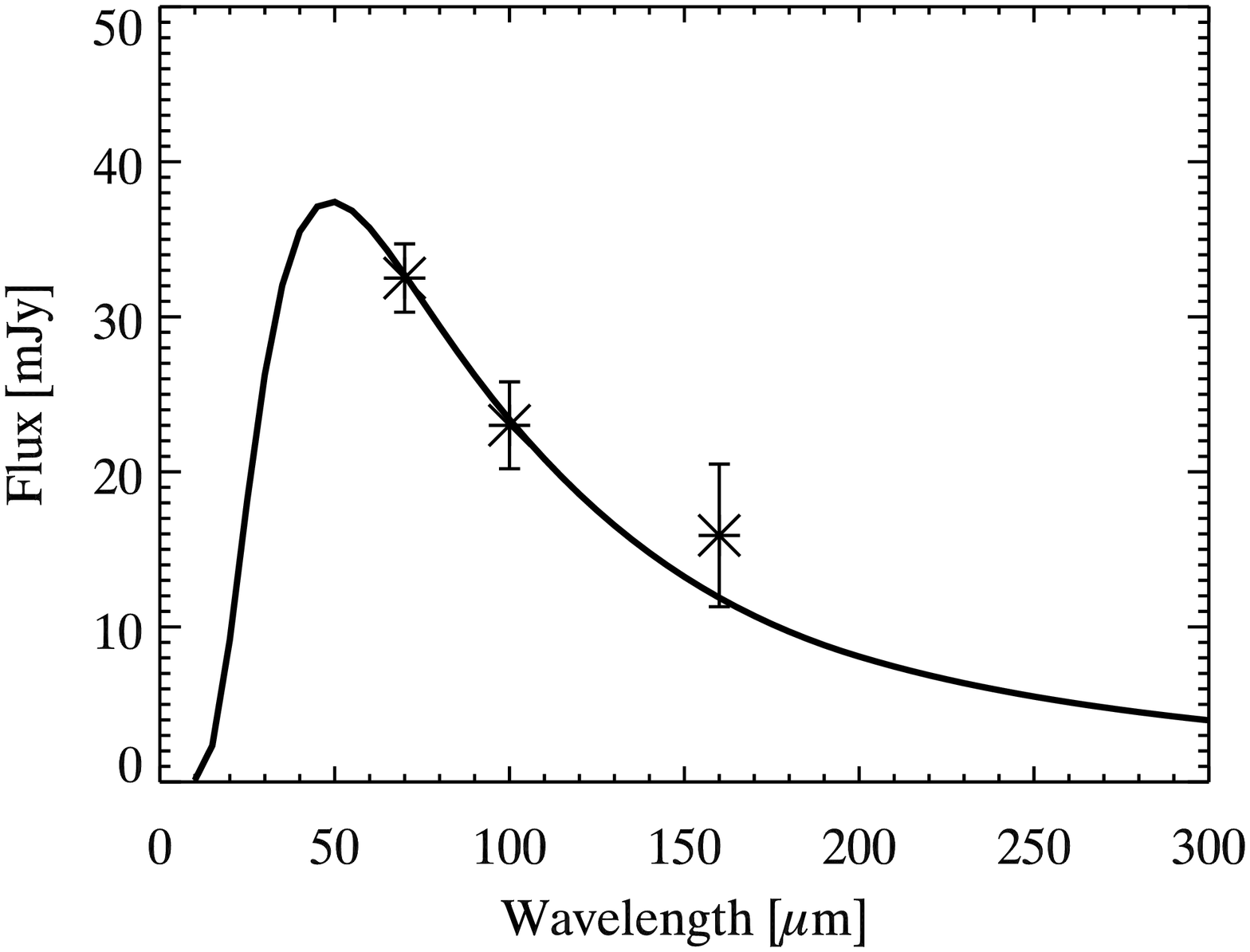}}%
\resizebox{!}{46mm}{\includegraphics[angle=90]{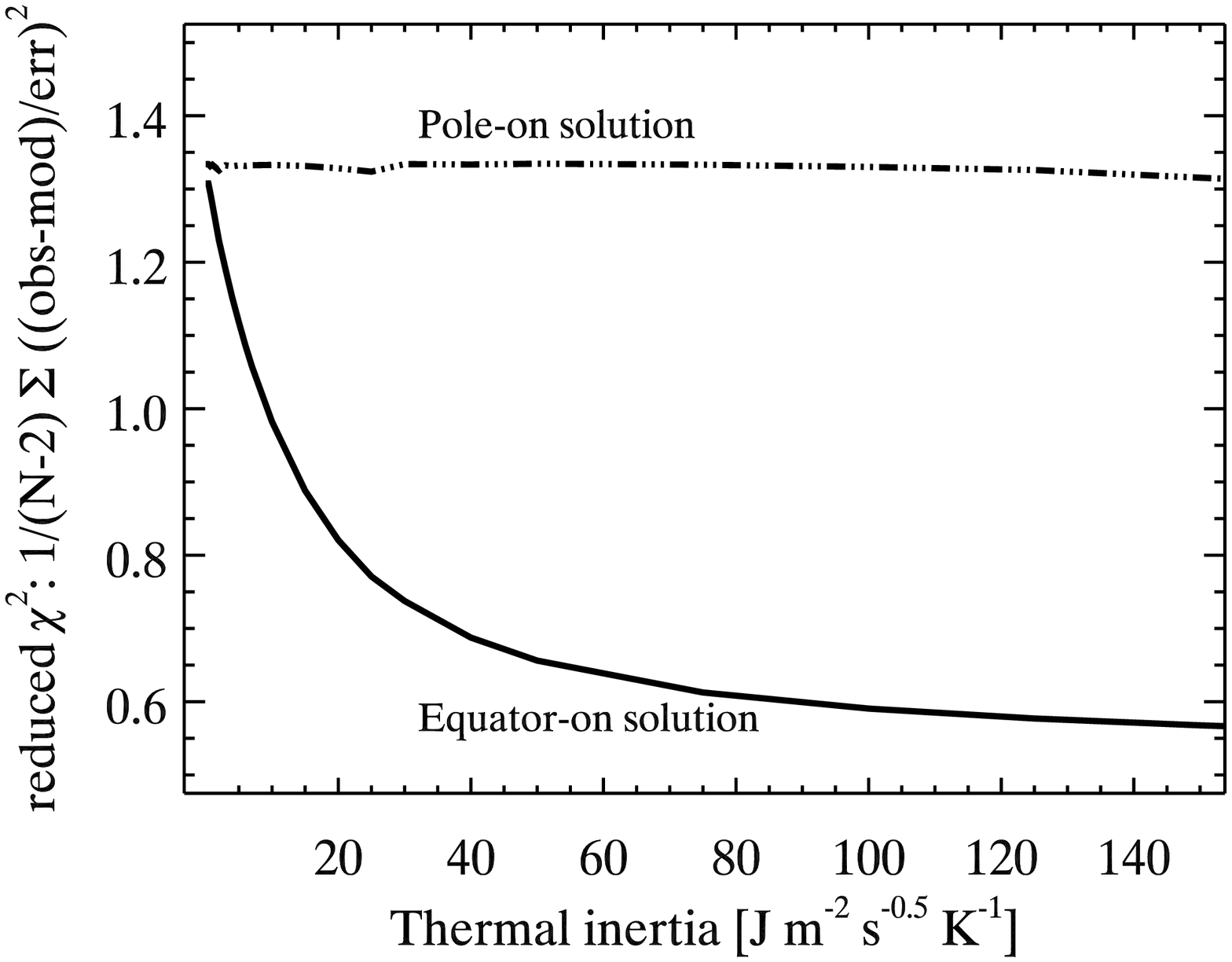}}
\end{center}\vspace*{-4mm}
\caption{Spectral energy distribution of 
\az{} in the far-infrared region, based on
Herschel/PACS measurements. Left panel: far-infrared measurements superimposed
are the best-fit NEATM curves with their respective uncertainty. 
Middle panel:
TPM model curve for thermal inertia of $100\,\tiunit$, rotation period
of $9.39\,{\rm h}$ and equator-on geometry. Right panel: the value of 
$\chi^2$ as the function of thermal inertia for pole-on and 
equator-on geometries. See text for further details.}
\label{fig:az60sed}
\end{figure*}

\subsection{Reflectance spectrum}
\label{sec:reflectance}

In order to accurately compare the surface colour characteristics of \az{} with
other TNOs \citep[see][]{lacerda2014},
we obtained a low resolution spectrum using the Optical System
for Imaging and Low Resolution Integrated Spectroscopy (OSIRIS) camera
spectrograph \citep{cepa2000,cepa2010} at the 10.4m Gran Telescopio
Canarias (GTC), located at the El Roque de los Muchachos Observatory (ORM) in
La Palma, Canary Islands, Spain. The OSIRIS instrument consists of a mosaic of
two Marconi CCD detectors, each with $2048\times4096$ pixels and a total
unvignetted field of view of $7.8^{\prime}\times7.8^{\prime}$, 
giving a plate scale of $0.127^{\prime\prime}/{\rm pixel}$.
However, to increase the signal to noise for our observations we
selected the $2\times2$ binning mode with a readout speed of 
$200\,{\rm kHz}$ (that has a gain of $0.95\,{\rm e^{-}/ADU}$ 
and a readout noise of $4.5\,{\rm e^-}$), as corresponds with the
standard operation mode of the instrument. A $300\,{\rm s}$ 
exposure time spectrum was obtained on January 28.17 (UTC), 2014 at an 
airmass of $X=1.14$ using the OSIRIS
R300R grism that produces a dispersion of $7.74\,{\rm \AA/pixel}$, 
covering the $4800-9000\,{\rm\AA}$ spectral range. A $1.5^{\prime\prime}$ slit 
width was used oriented at the parallactic angle.

Spectroscopic reduction has been done using the standard IRAF tasks. Images
were initially bias and flat-field corrected, using lamp flats from the GTC
Instrument Calibration Module. The two-dimensional spectra were then wavelength
calibrated using Xe+Ne+HgAr lamps. After the wavelength calibration, sky
background was subtracted and a one dimensional spectrum was extracted. To
correct for telluric absorption and to obtain the relative reflectance, G2V
star Land102\_1081 \citep{landolt1992} was observed using the same spectral
configuration and at a similar airmass immediately after the Centaur
observation. The spectrum of the \az{} was then divided by that of
Land102\_1081, and then normalized to unity at $0.55\,{\rm\mu m}$.

The derived spectrum is displayed in Fig.~\ref{fig:reflectance}.
Based on this spectrum, the slope parameter of this object is found to
be $S^\prime=13.4\pm3.0\,\%(1000\mathrm{\AA})^{-1}$ by a 
linear fit across the interval $5000-9000\,{\rm\AA}$. 

The measured photometric colours ($g'-r'=0.755\pm0.018$ and 
$0.72\pm0.05$ on 2013.11.04 and 2014.01.28, respectively) are in 
complete accordance with the derived spectral slope. 
The spectrum was normalized at $5500\,{\rm \AA}$, just between the
$g'$ and $r'$ bands. We can therefore write $S^\prime$ to equation (2)
of \cite{jewitt2002}, if we write SDSS colours instead of Bessel ones, and set
$\Delta\lambda=1480\,{\rm \AA}$. This results in a synthetic colour index from
spectral slope $(g-r)_{synth}=0.76\pm0.04$, in a perfect agreement with our
photometry within the errors.

\subsection{Near-infrared photometry}
\label{sec:nirphoto}

CCD observations of \az{} were obtained on 24 September 2013 with
the 4.2-m William Herschel Telescope at La Palma Observatory,
equipped with the LIRIS instrument. LIRIS is a near-IR
imager/spectrograph, which uses a $1{\rm k}\times1{\rm k}$ 
HAWAII detector with a field of view of $4.27^\prime\times4.27^\prime$. 
The number of exposures taken in different filters are: $5\times30\,{\rm s}$ 
in $Y$ and $J$, $15\times20\,{\rm s}$ in $H$, $15\times13\,{\rm s}$ in 
CH4, and $180\times20\,{\rm s}$ exposures in $K_{\rm s}$. 
Local comparison stars were selected from the 2MASS
catalogue and magnitude transformation were applied following \cite{hodgkin2009}.

The result of the photometry is $Y=18.66\pm0.08$, $J=18.34\pm0.05$, 
$H=18.00\pm0.06$ and $K_{\rm s}=17.72\pm0.10$ 
where $Y$ refers to the UKIDDS system, while $JHK_{\rm s}$ are 2MASS
magnitudes. Thus, \az{} exhibits almost exactly solar colour
indices, with a slightly redder slope than solar. Namely
$Y-J=0.32$, $J-H=0.34\pm0.07$ and $H-K_s=0.28\pm0.11$ while according to
\cite{casagrande2012} and estimating solar $Y-J$ according to \cite{hodgkin2009},
the respective solar colours are   $(Y-J)_\odot=0.235\pm0.018$
$(J-H)_\odot=0.286\pm0.018$ and $(H-K_s)_\odot=0.076\pm0.018$.
We note here that LIRIS is equipped with Mauna Kea Observatories
(MKO) system of $J$, $H$ and $K$ filters. According to 
\cite{hodgkin2009}\footnote{See their equations (6), (7) and (8)}, the 
expected systematic differences between 2MASS and LIRIS/MKO colours of
\az{} is in the range of $-0.010 \dots +0.015$, which is definitely smaller
than the photometric uncertainties.
This observation indicates a flat and featureless spectrum of \az{}:
the slope is equivalent in the infrared and in the optical, being quite 
similar to dormant cometary nuclei.

During the observation the heliocentric and geocentric distance of
\az{} were $8.87$ and $8.32\,{\rm AU}$, respectively, indicating an
absolute mid-IR brightness of $J=9.00\pm0.06$ without correcting
for the solar phase angle.


\begin{table}
\caption{Orbital and optical data for 
\az{} at the time of the Herschel 
observations.} 
\label{table:auxdata}
\begin{center}\begin{tabular}{llr}
\hline
Quantity			& Symbol 	& Value \\
\hline
Heliocentric distance		& $r$           & $8.702$\,AU           \\
Distance from Herschel		& $\Delta$      & $8.560$\,AU           \\
Phase angle			& $\alpha$      & $6.\!\!^\circ6$       \\
Absolute visual magnitude	& $H_V$		& $10.45\pm 0.10$       \\
\hline
\end{tabular}\end{center}\vspace*{-3mm}
\end{table}

\section{Thermal emission modelling}
\label{sec:thermal}

\subsection{Near-Earth Asteroid Thermal Model}

The basic physical properties such as albedo and diameter can be 
obtained by combining optical brightness data with thermal emission.
Assuming an absolute optical brightness for a certain object, the
higher the thermal emission, the smaller the actual albedo and hence
the larger the diameter. 
Due to the observing strategies constrained by the spatial attitude of the
Herschel, objects close to the ecliptic are likely observable by Herschel 
during quadratures. Nevertheless, in quadratures these minor objects
exhibit a large phase angle, hence Standard Thermal 
Model \citep[STM][]{lebofsky1986} might not be as accurate as it
is desired. \az{} had a phase angle of $\alpha=6.6^\circ$ at the time 
of our Herschel/PACS observations. In order to have accurate estimates 
for larger phase angles, we employed the Near-Earth Asteroid Thermal
Model \citep[NEATM][]{harris1998}: this model integrates the thermal emission
for arbitrary viewing angles. Throughout the modelling we use the 
heliocentric and geocentric distances at the instance of the Herschel/PACS
measurements, namely $r_{\rm helio}=8.702\,{\rm AU}$ 
and $r_{\rm geo}=8.560\,{\rm AU}$.

The diameter and albedo can then be derived in a similar manner like in
our earlier works \citep[see e.g.][]{kiss2013,pal2012sedna}. As an input
for the fit procedures, we used the previously obtained thermal fluxes
(see Table~\ref{table:herscheldata}) and the absolute brightness
$H_V=10.45\pm0.10$ (derived earlier, see above). 

The absolute physical parameters (diameter, albedo and beaming parameter)
have been obtained in a Monte-Carlo fashion. In each step, a Gaussian 
value were drawn for the four input values (three thermal fluxes and the
absolute brightness $H_V$) and the model parameters were adjusted via a 
nonlinear Levenberg-Marquardt fit. A sufficiently long series of such steps 
yields the best fit values as well as the respective uncertainties and
correlations. 
This procedure were performed in two iterations. First, we let the value 
for the beaming parameter $\eta$ to be floated. This run resulted relatively
high correlations between the parameters and a highly long-tailed
distribution for $\eta$: we found that the mode for $\eta$ was $0.8$ while 
the median is $2.6$ and the uncertainties yielded by the lower and upper 
quartiles are $2.6^{+2.9}_{-1.1}$. This skewed distribution 
is due to the fact that beaming parameters cannot
really be constrained if thermal fluxes are not known for shorter wavelengths
(i.e. shorter than the peak of the spectral energy distribution). 
Hence, in the next run we used $\eta$ as 
an input (instead of an adjusted variable) while its value was drawn 
uniformly between $0.8$ and $2.6$. This domain is also in accordance with 
the possible physical domain of the beaming parameter 
\citep[see also Fig. 4 of][]{lellouch2013}.
The results of this second run were 
$d=62.3\pm5.3\,{\rm km}$, $p_V=0.029\pm0.006$ while the beaming parameter
can be written as $\eta=1.7\pm0.9$. The resulting albedo refers to a remarkably
dark surface. The fluxes along with the best-fit NEATM model curve
are shown in the left panel of Fig.~\ref{fig:az60sed}.

\subsection{Thermophysical Model}

In addition to the derivation of the NEATM parameters, we conducted
an analysis of thermal emission based on the asteroid thermophysical model
\citep[TPM, see][]{muller1998,muller2002}. The observational
constraints employed by this model was the thermal fluxes
(see Table.~\ref{table:herscheldata}), the absolute magnitude of 
$H_V=10.45\pm0.10$ (see earlier), the rotational period of $9.39\,{\rm h}$
as well as the actual geometry at the time of Herschel observations
(see the values for phase angle, heliocentric and geocentric distances above).

\refmark{Our procedures have shown that the best-fit model occurs at high
thermal inertia values. Assuming an equator-on geometry, a value for 
reduced $\chi^2 \lesssim 1$ corresponds to $\Gamma \gtrsim 10\,\tiunit$ (see also 
the right panel of Fig.~\ref{fig:az60sed}), however, the gradually decreasing
form of the function $\chi^2(\Gamma)$ implies a lower limit of 
$\Gamma \gtrsim50\,\tiunit$. The corresponding values at 
$\Gamma=50\,\tiunit$ for geometric albedo and diameter 
are $p_V=0.028$ and $d=64.9\,{\rm km}$, respectively.}
These values are also compatible within uncertainties with the ones
derived from NEATM analysis (see above). 
The spectral energy distribution provided by these TPM values 
are shown in the middle panel of Fig.~\ref{fig:az60sed}. This value for the 
thermal inertia is close to the values of $30-300\,\tiunit$ reported 
for comets \citep[see e.g.]{julian2000,campins2000,davidsson2013}
as well as the value of $10-50\,\tiunit$ for 
67P/Churyumov-Gerasimenko \citep{gulkis2015}. We note here 
that models either with thermal inertia values smaller than $50\,\tiunit$ 
or having an assumption for pole-on view underestimate 
the observed flux at $160\,{\rm\mu m}$.

\refmark{Our findings for large preferred values of the beaming parameter $\eta$
as well as for the thermal inertia $\Gamma$ (even $\Gamma \gtrsim 10\,\tiunit$)
can be compared with the statistical expectations of \cite{lellouch2013}. }
By considering the small heliocentric distance of this object, both of 
these values are expected to be smaller \citep[see Figs. 6 and 11 in][]{lellouch2013}.


\section{The dynamics of \az{}}
\label{sec:dynamics}

\az{} moves on a highly eccentric orbit, with perihelion between the 
orbits of Jupiter and Saturn. The best fit solution for the epoch of 
March 15, 2015 is shown in Table~\ref{table:orbit}.

\begin{table}
\caption[]{The best-fit orbital solution (semi-major axis $a$, perihelion 
distance $q$, eccentricity $e$, inclination $i$, longitude of ascending 
node $\Omega$, argument of perihelion $\omega$, mean anomaly $M$ and
perihelion date $T_{\rm peri}$),
and associated uncertainties, for \az{}, taken from the 
Minor Planet Center (http://www.minorplanetcenter.net) on March 15, 2015.
Note that the uncertainties of the orbital elements involved throughout 
the planning of the observations significantly larger than these due to the
shorter arcs available at that time.}
\label{table:orbit}
\begin{center}
\begin{tabular}{cc}
\hline
 a (AU)		& 829.7				\\
 q (AU)		&    7.908098 	$\pm$ 0.000014	 	\\
 e	        &    0.990468 	$\pm$ 0.000010	 	\\
 i (deg)    	&   16.535760 	$\pm$ 0.000011 		\\
$\Omega$ (deg)	&  349.21122 	$\pm$ 0.00002 		\\
$\omega$ (deg)  &  158.14327 	$\pm$ 0.00021 		\\
M (deg) 	&    0.00876  				\\
$T_{\rm peri}$  & 2456988.0641  $\pm$ 0.0032		\\
Epoch (JD) 	& 2457200.5  				\\
\hline
\end{tabular}
\end{center}
\end{table}

In order to assess the dynamical history, and potential future behaviour of 
\az{}, we follow a well-established route 
\citep[see e.g.][]{horner2004a,horner2004b,horner2010,horner2012,kiss2013}, 
and used the Hybrid integrator within the n-body dynamics package 
MERCURY \citep{chambers1999}, to follow the evolution of a swarm of test particles 
centered on the best-fit orbit for the object, in order to get a statistical 
overview of the object's behaviour. A total of 91,125 test particles were 
created, distributed uniformly across the region of orbital element 
phase space within $\pm3\sigma$of the best-fit perihelion distance, 
$q$, eccentricity, $e$, and inclination, $i$. In this manner, we 
created a grid of $45\times45\times45$ test particles distributed 
in even steps across the $\pm3\sigma$ error ranges about the nominal 
best fit orbit in each of the three orbital elements studied. 
Each of these test particles was then followed in our integrations, 
with its orbit evolving under the gravitational influence of the giant 
planets Jupiter, Saturn, Uranus and Neptune, for a period of four 
billion years. Test particles were considered to have been ejected 
from the Solar System (and were therefore removed from the integrations) 
if they reached a barycentric distance of 10,000\,AU. 
Similarly, any test particles that collided with one of the giant 
planets, or with the Sun, were removed from the simulations. 
Each time a test particle was removed in either of these manners, 
the time at which the removal occurred was recorded, allowing us to track 
the number of surviving test particles as a function of time. The results 
of our simulations are shown below, in 
Figs~\ref{fig:clones}~and~\ref{fig:stability}.
 
\begin{figure}
\begin{center}
\includegraphics[width=88mm]{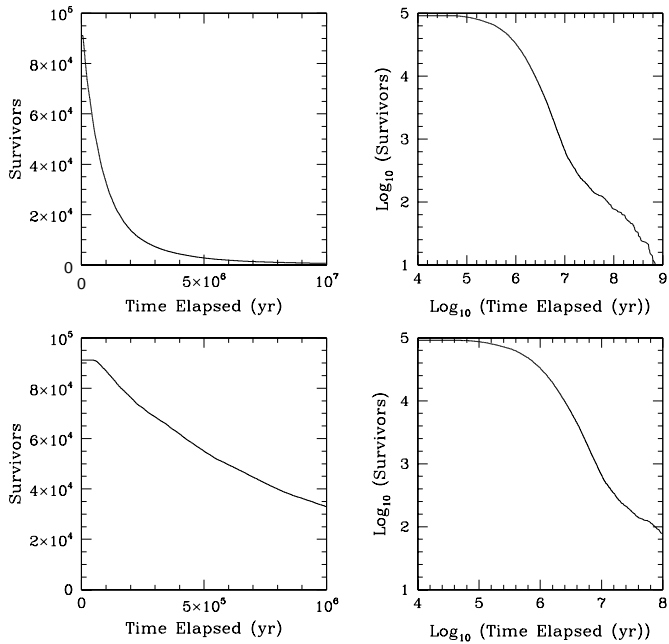}
\end{center}
\caption[]{The decay of our population of 91,125 clones of 
\az{} as a function of the time elapsed in our integrations. 
The plots on the right show the same data as those on the left, 
but are plotted on a log/log scale. The upper panels show the decay 
of the population over the first billion years of the four billion 
year integrations, whilst the lower panels show the decay over the 
first million years.}
\label{fig:clones}
\end{figure}
 
It is immediately apparent that the population of clones of \az{} is highly 
dynamically unstable, with 63.9\% of the particles (58191 of 91125) being 
removed from the simulations within the first million years of the 
integrations, as a result of either ejection or collision with one of the 
giant planets or the Sun. Half of the test particles are ejected within 
the first 682\,kyr of the integrations, revealing that the orbit of 
\az{} is more than two orders of magnitude more unstable than that of 
the similar object \dr{} \citep{kiss2013}. 
  
\begin{figure}
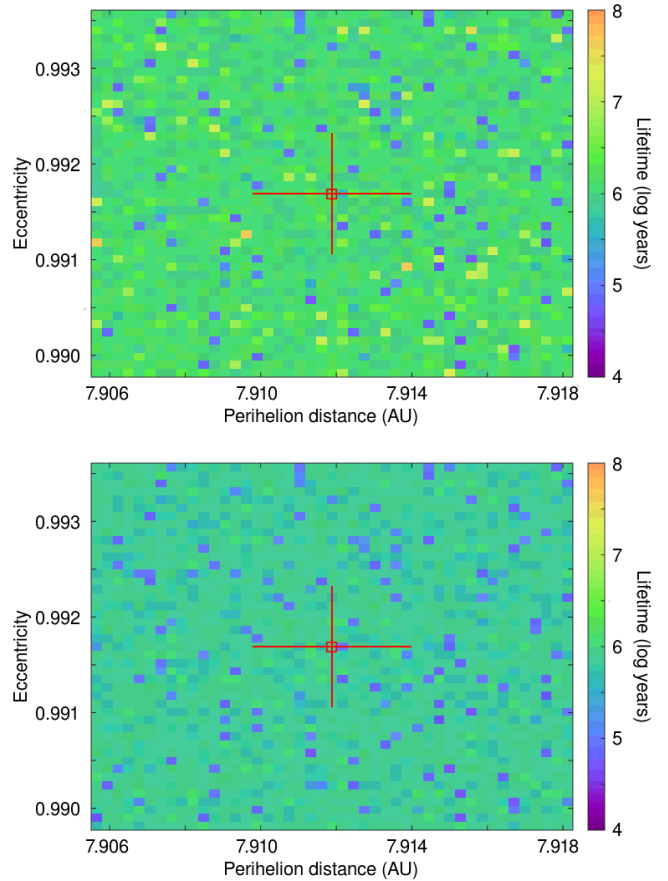

\begin{center}
\includegraphics[width=88mm]{az60-meanlifetime}
\includegraphics[width=88mm]{az60-medianlifetime}
\end{center}
\caption[]{The mean (upper) and median (lower) lifetimes of 
\az{}, as a function of the initial perihelion distance, q, and 
eccentricity, e, of the orbit tested. The location of the best-fit 
orbital solution for \az{}, as detailed in Table~\ref{table:orbit}, 
is shown by the hollow square at the center of the figure, 
with the $\pm$1$\sigma$ uncertainties on the perihelion distance and 
eccentricity denoted by the solid black lines that radiate from that box. 
Each coloured square in the figures shows the mean (or median) lifetime of 
the 45 individual runs carried out at that particular a-e location. 
Each of those 45 runs tested a different orbital inclination for 
\az{}, evenly distributed across the $\pm3\sigma$ uncertainty range 
on the best-fit orbital solution. As was the case with the 
high-eccentricity Centaur \dr{} \citep{kiss2013}, the stability 
of the orbit of \az{} does not vary significantly across the range 
of perihelion distance and eccentricities tested in this 
work -- a reflection of the relatively high precision with 
which the object's orbit is known.}
\label{fig:stability}
\end{figure}

The orbit of \az{} proves to be highly dynamically unstable on timescales of 
just a few hundred thousand years. Fully half of the test particles in our 
simulations were removed from the simulations within just 682\,kyr, and 
almost two-thirds were removed within the first million years. This extreme 
level of instability is not, however, that surprising -- \az{} passes 
through the descending node of its orbit\footnote{as can be seen 
in the elegant Java visualization of the object's orbit at 
http://ssd.jpl.nasa.gov/sbdb.cgi? sstr=2013\%20AZ60;orb=1;cov=0;log=0;cad=0\#orb} 
at essentially the same time it passes through perihelion, 
maximizing the likelihood that it will be perturbed by either Jupiter 
or Saturn. This extreme level of instability is typical of objects 
moving on Centaur like orbits \citep[e.g.][]{horner2004a,horner2004b} 
and suggests that \az{} may only recently have been captured to its 
current orbit. This argument is supported by the fact that, averaged 
over our entire population of 91,125 test particles, the mean 
lifetime of \az{} is just 1.56\,Myr.

Given that \az{} exhibits such extreme instability, and may well be 
relatively pristine object, it is interesting to consider whether it 
will have experienced significant solar heating, and cometary activity, 
over its past history. As a result of our large dynamical dataset on the 
evolution of \az{}, it is possible to determine the fraction of the 
population of clones that may one day evolve onto Earth-crossing 
orbits, and the fraction of the population that approach the Sun to 
within a given heliocentric distance at some point in their lifetime. 
Since dynamical evolution under the influence of gravity alone is a 
time-reversible process, we can use these values to estimate the probability 
that \az{} has moved on orbits that bring it within those heliocentric 
distances at some point in the past, before being ejected to its 
current orbit. Due to the extreme instability exhibited by \az{}, 
we found that a relatively small number of the total population of 
clones were captured to Earth-crossing orbits through their 
lifetimes. Indeed, just 3805 of the 91125 test particles we 
studied (just 4.2\% of the population) became Earth-crossing 
at any point in our integrations, and the total fraction of 
the object's lifetime spent as an Earth-crossing object (averaged across 
all 91125 test particles) was 0.12\%. Our results for a variety of 
other perihelion distances are displayed in Table~\ref{table:evolution}, 
together with estimates of the mean amount of time for which clones of 
\az{} exhibited perihelion distances smaller than the specified value.

\begin{table}
\begin{tabular}{ccc}
\hline
                      	&     Number of clones	& Percentage of total 	\\
                      	&                       & integration time  	\\
\hline
Earth-crossing          & 	3805	& 0.118 \\ 
($q<1.0616\,{\rm AU}$)	\\ 
$q<2\,{\rm AU}$ 	& 	6005	& 0.291 \\
$q<4\,{\rm AU}$ 	&	12272 	& 0.329 \\
$q<6\,{\rm AU}$ 	& 	27150	& 2.06 \\
\hline
\end{tabular}
\caption[]{The number of the 91,125 
clones of \az{} simulated in this work that evolved to orbits 
with perihelion distances smaller than 2, 4, and 6\,AU, and the 
number that evolved onto Earth-crossing orbits 
(following Horner et al., 2003). For each of these values, 
we also give the fraction of the total integration time, 
across all 91,125 clones, for which clones have perihelion 
distances within these limits. We note that this is the 
fraction of the time for which the perihelion distance was 
less than the stated amount and not the fraction of time the 
clones spend within that heliocentric distance. Even when moving on 
an orbit with perihelion within that of the Earth, a given clone will 
spend the vast majority of its time beyond that distance, and only a 
tiny fraction within it.}
\label{table:evolution}
\end{table}

\begin{figure}
\begin{center}
\resizebox{85mm}{!}{\includegraphics{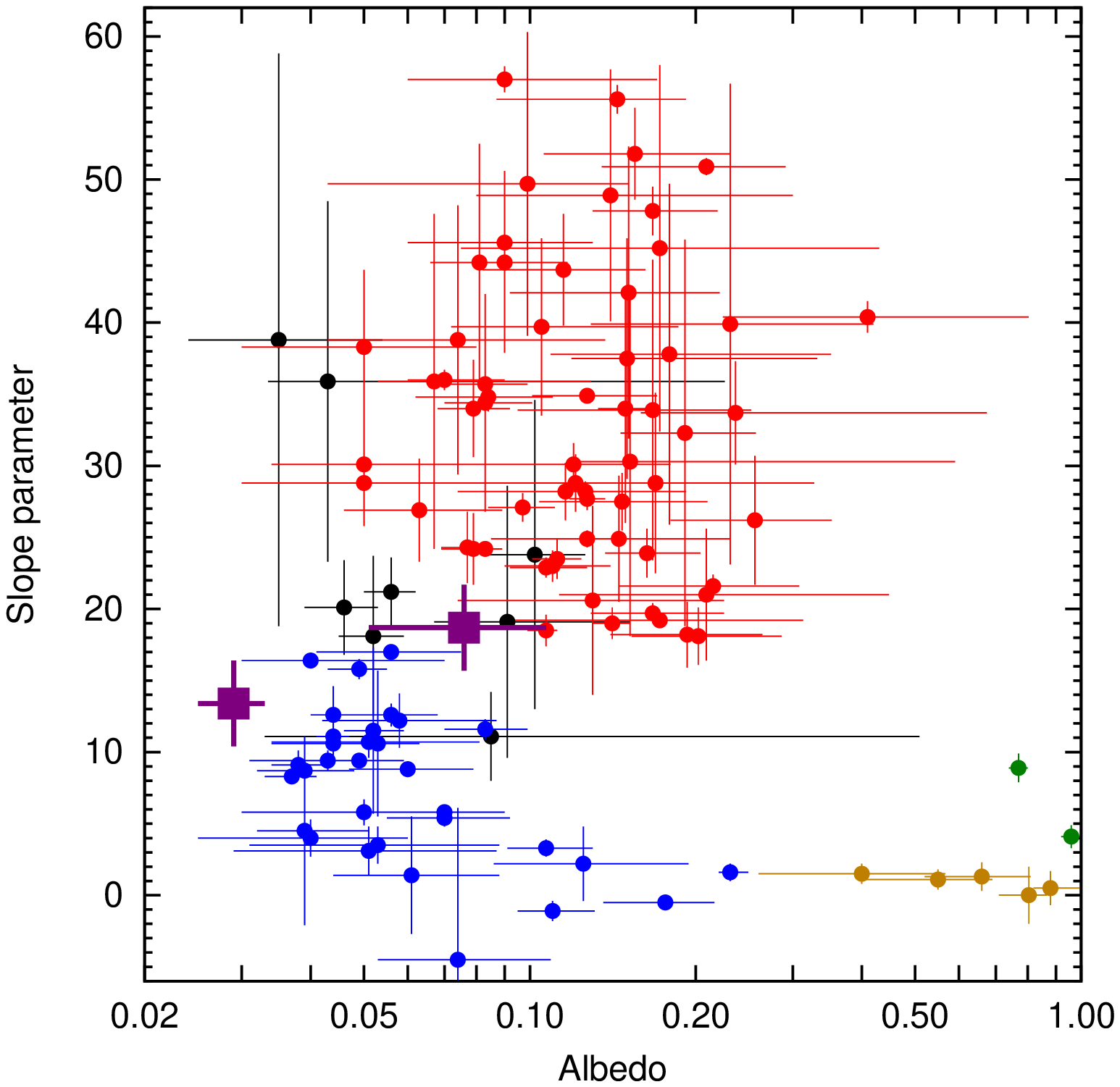}}
\end{center}\vspace*{-4mm}
\caption{Slope parameter vs. albedo relations for 111 TNOs, including
\az{} and \dr{}. Data (except for these two latter objects)
have been taken from \cite{lacerda2014}. The blue
and red dots indicate the two major groups identified by \cite{lacerda2014},
black points represent ambiguous objects (due to their
large respective uncertainties) while green and yellow dots show the 
large bodies and Haumea-type surfaces, respectively.
The isolated purple square shows the place of \az{} at the very left
side of the diagram. The other purple square indicates \dr{},
just in between of the  dark neutral (blue) and  bright red (red) object
groups.}
\label{fig:albedoslope}
\end{figure}

\section{Discussion}
\label{sec:discussion}

Since the orbit of \az{} is highly eccentric, and takes the object out 
to approximately $1950\,{\rm AU}$, it is clear that it spends the vast 
majority of its orbit at large heliocentric distance. It is quite 
plausible that \az{} is a relatively recent entrant to the inner 
Solar System. Hence, it is interesting to consider how much time, cumulative 
over its entire history since it was first emplaced on a planet 
crossing orbit, has it spent at a heliocentric distance 
of less than 1, 10 or 100\,AU. Again, we can take advantage of 
the large dynamical dataset available to us from our integrations to 
get a feel for the amount of time the object will have spent within 
these distances. Clearly, this is only an estimate (and implicitly assumes 
that, prior to its injection to a planet-crossing orbit, the object was 
well beyond the 100\,AU boundary -- i.e. that it was injected from the 
inner or outer Oort cloud, rather than the trans-Neptunian region). 
Given that implicit assumption, we find that, on average, clones 
of \az{} spend just 6.68 years within 1\,AU of the Sun, 4620 years 
within 10\,AU of the Sun, and 273,000 years within 100\,AU of the Sun. 
The time spent within 10 and 100\,AU is strongly biased by a few particularly 
long-lived clones, especially those that are captured onto Centaur-like orbits. 
We note that more than two-thirds of the clones (64904 objects) spent less 
than a thousand years within 100\,AU of the Sun, and 37493 (41.1\%) 
spent less than one hundred years within 100\,AU. Taken into considerations, 
our dynamical results suggest that \az{} has only recently been 
captured to its current planet-crossing orbit, and that it is 
quite likely that it is a relatively pristine object. Indeed, it 
seems highly probable that the surface of \az{} has experienced 
only minimal outgassing and loss of volatiles since being captured to 
a planet crossing orbit, and so it represents a particularly interesting 
object to target with further observations as it pulls away from the 
Sun following its recent perihelion passage.

Outer Solar System objects can also be characterized in a way
recently put forward by \cite{lacerda2014}, using their visual range
colours and albedos. In this frame, Centaurs and trans-Neptunian objects form
typically two clusters, a dark-neutral and a bright-red one 
\citep[see Fig. 2 in][]{lacerda2014}. In this scheme, \az{} is located at the
dark (very low albedo) edge of the dark-neutral cluster, see
Fig.~\ref{fig:albedoslope}. \az{} is even darker than the object
\gz{}, the object with the lowest albedo in the sample of \cite{duffard2014}.
Objects with characteristics similar to our target belong rather to ``dead 
comets'' or Jupiter family comets which are the end states of Centaurs and Oort cloud 
comets \citep[Fig. 4 in][]{lacerda2014}; in this sense \az{} is more similar 
to objects in the inner Solar System than those in the trans-Neptunian 
population. We also checked the distribution of the slope parameters of
various Centaurs based on \cite{fornasier2009}. Although in that work,
a correlation between the slope parameters and orbital eccentricity were
suspected (the higher the eccentricity, the redder the object), 
the large eccentricity of \az{} do not fit in this model since it has 
definitely lower slope parameter than the mean of that sample of Centaurs. 

While the dynamical analysis indicate that \az{} has recently been
pulled from the Oort cloud, in the case of this object there is a
much higher likelihood that it has spent a considerable time in the
inner Solar System then e.g. in the case of \dr{}, which might
just be in a transitional phase between the two main albedo-colour
clusters \citep{kiss2013}.


\begin{acknowledgements}
We thank the comments and the thoughtful review of the anonymous referee. 
The work of A.~P., Cs.~K. and R.~Sz. has been supported by 
the grant LP2012-31 of the Hungarian Academy of Sciences
as well as the ESA PECS grant No. 4000109997/13/NL/KML of the 
Hungarian Space Office and the European Space Agency, and the K-104607 
and K-109276 grants of the Hungarian Research Fund (OTKA). The work of Gy.~M.~Sz.
has also been supported by the Bolyai Research Fellowship of the 
Hungarian Academy of Sciences. Additionally, Gy.~M.~Sz. and K.~S. has been supported by
ESA PECS No. 4000110889/14/NL/NDe and the City of Szombathely under 
agreements No. S-11-1027 and 61.360-22/2013. K.~S. has also been supported by 
the ``Lend\"ulet'' 2009 program of the Hungarian Academy of Sciences.
J.~L. acknowledge support from the project AYA2012-39115-C03-03 
(MINECO, Spanish Ministry of Economy and Competitiveness).
Part of this work was supported by the German DLR project number 50~OR~1108.
Based on observations made with the Gran Telescopio Canarias (GTC), instaled in
the Spanish Observatorio del Roque de los Muchachos (ORM) of the Instituto de
Astrof\'{\i}sica de Canarias (IAC), in the island of La Palma, the William 
Herschel telescopes (WHT) operated in the ORM by the Isaac Newton Group and on
observations made with the IAC-80 telescope operated on the island of Tenerife
by the IAC in the Spanish Observatorio del Teide. WHT/LIRIS observations
were carried out under the proposal SW2013a15.

\end{acknowledgements}


{}

\end{document}